# Block Designs and *K*-Geodetic Graphs: A Survey

**Frasser C.E.**

*Ph.D. in Engineering Sciences, Odessa National Polytechnic University, Ukraine*

**Abstract**

*K*-geodetic graphs (*K* capital) are defined as graphs in which each pair of nonadjacent vertices has at most *K* paths of minimum length between them. A *K*-geodetic graph is geodetic if $K = 1$, bigeodetic if $K = 2$ and trigeodetic if $K = 3$. *K*-geodetic graphs are applied effectively to the solution of several practical problems in distinct areas of computer science, hence the importance of their study. Four problems are central to the study of *K*-geodetic graphs, namely, characterization, construction, enumeration and classification. The problems of finding the general classification of *K*-geodetic graphs for each of their classes $K = 1, 2, 3$ are open. The present paper is a survey dedicated to the construction of *K*-geodetic graphs for $K = 1, 2, 3$ using balanced incomplete block designs (BIBDs). To this purpose, we use block designs as combinatorial structures defined in terms of completely predetermined parameters, which is essential for the easy construction of the *K*-geodetic graphs described in this survey.

## Introduction

A *balanced incomplete block design* (or simply a block design) on a set $S$ with $|S| = n$, where $|S|$ is the cardinal number of $S$, is a family of subsets $B_1, B_2, …, B_b$ of $S$ called blocks such that:
(a)  $|B_i| = k, 1 \le i \le b$.
(b)  If $x \in S$, then $x$ belongs to exactly $r$ blocks $B_i$.
(c)  If $x, y$ are distinct elements of $S$, then $\{x, y\}$ is contained in exactly $\lambda$ blocks.
This block design is denoted $(b, n, r, k, \lambda)$.

Block designs are defined in terms of certain parameters whose values determine the answer to the question of existence, that is, which values of these parameters produce the configuration in question and which do not. Given a $(b, n, r, k, \lambda)$-design, the following are necessary conditions that its parameters must satisfy:

$$bk = nr, \quad r(k - 1) = \lambda(n - 1). \tag{1}$$

The conditions previously described on the parameters of a block design are necessary, but not sufficient. It means that we can use them to rule out the existence of a block design for certain groups of parameters. However, being given the values of the parameters, which satisfy the conditions (1), does not guarantee the existence of a block design with those parameters. There are many groups of possible parameters for which the existence problem has not been settled.

A block design with $k = 3$ is called a *triple system*. A triple system with $\lambda = 1$ is called a *Steiner triple system*. A necessary condition for the existence of a Steiner triple system is that $n \equiv 1$ or $3 \pmod 6$. Steiner posed as a problem in 1853 whether these necessary conditions were sufficient for its existence. This was solved affirmatively by Reiss in 1859 [2]. The specific parameters of a triple system [1] can be written in the form:

$$\left(\frac{\lambda n(n-1)}{6}, n, \frac{\lambda(n-1)}{2}, 3, \lambda\right) \text{ for a fixed } \lambda = 1, 2 \text{ or } 3. \tag{2}$$

**Theorem 1.** (*Hanani* [2]) *A necessary and sufficient condition for the existence of a balanced incomplete block design* $\left(\frac{\lambda n(n-1)}{6}, n, \frac{\lambda(n-1)}{2}, 3, \lambda\right)$ *and any* $\lambda$ *is that*

$$\lambda n(n-1) \equiv 0 (\mathrm{mod}\ 6) \quad \text{and} \quad \lambda(n-1) \equiv 0 (\mathrm{mod}\ 2), \tag{3}$$

*which is obvious from the parameters of the block design considering that* $\frac{\lambda n(n-1)}{6}$ *and* $\frac{\lambda(n-1)}{2}$ *are positive integers. Conditions* (3) *can be restated in the form*:

$$\begin{array}{ll} \text{If } \lambda \equiv 1 \text{ or } 5 (\mathrm{mod}\ 6), & n \equiv 1 \text{ or } 3 (\mathrm{mod}\ 6); \\ \text{If } \lambda \equiv 2 \text{ or } 4 (\mathrm{mod}\ 6), & n \equiv 0 \text{ or } 1 (\mathrm{mod}\ 3); \\ \text{If } \lambda \equiv 3 (\mathrm{mod}\ 6), & n \equiv 1 (\mathrm{mod}\ 2); \\ \text{If } \lambda \equiv 0 (\mathrm{mod}\ 6), & \text{no restrictions on } n. \end{array} \tag{4} \quad \square$$

A design with $b = n$ is called symmetric. In such a design $r = k$ and hence such structure is called $(b, k, \lambda)$-design. For symmetric designs, there is an additional restriction for their existence.

**Theorem 2.** (*Bruck − Ryser − Chowla* [2]) *Let* $b$, $k$, $\lambda$ *be integers for which there exists a symmetric* $(b, k, \lambda)$-*design. If* $b$ *is even, then* $k - \lambda$ *equals a perfect square. If* $b$ *is odd, then the equation*

$$x^2 = (k - \lambda)y^2 + (-1)^{(b-1)/2}\lambda z^2 \tag{5}$$

*has a solution in integers* $x, y, z$ *not all zero.* $\qquad \square$

The specific parameters of a symmetric block design [1] can be written in the form:

$$\left(\frac{n^2 + n + \lambda}{\lambda}, n + 1, \lambda\right) \text{ for a fixed } \lambda = 1, 2 \text{ or } 3. \tag{6}$$

**Theorem 3.** *For a symmetric* $\left(\frac{n^2 + n + \lambda}{\lambda}, n + 1, \lambda\right)$-*design, the following is true*:

(a) *If* $\lambda = 1$, *then a symmetric* $(n^2 + n + 1,\ n + 1, 1)$-*design does exist for* $n$ *a prime power* [2].

(b) *If* $\lambda = 2$, *then a symmetric* $\left(\frac{n^2 + n + 2}{2}, n + 1, 2\right)$-*design does exist for* $n \equiv 1$ *or* $2 (\mathrm{mod}\ 4)$, $2 \le n \le 10$, $n - 1$ *a perfect square or* $n \equiv 0$ *or* $3 (\mathrm{mod}\ 4)$, $3 \le n \le 12$, $n - 1$ *a prime power* [1].

(c) *If* $\lambda = 3$, *then a symmetric* $\left(\frac{n^2 + n + 3}{3}, n + 1, 3\right)$-*design does exist for* $n \equiv 0$ *or* $2 (\mathrm{mod}\ 3)$, $3 \le n \le 14$, $n \ne 12$ [1]. $\qquad \square$

The set of vertices and edges of a graph $G$ are denoted $V(G)$ and $E(G)$, respectively. An *undirected graph* is one having edges with no direction. Two vertices of a graph $G$ are *adjacent* if they are connected by an edge. A graph in which every pair of its vertices is adjacent is called *complete*. The complete graph on $n$ vertices is denoted $K_n$. A *path* in a graph $G$ from vertex $v_0$ to vertex $v_n$ is a sequence $v_0 v_1 \ldots v_n$ of different vertices and is denoted $P(v_0, v_n)$. The number of edges of a path $P$ in $G$ determines the length of this path and is represented by $|P|$. The length of a shortest path connecting vertices $u$ and $v$ in $G$ represents the *distance* between these two vertices. The greatest distance between any pair of vertices of $G$ is called the *diameter* of $G$, which is denoted $d(G)$. The number of edges incident to $v$ is called the *degree* of vertex $v$ and is denoted $deg\ (v)$. $G$ is said to be *regular of degree k* (or *k-regular*) if every vertex of $G$ has equal degree $k$ and *biregular* with *degree sequence* $(l, k)$ if for any vertex $v$ of $G$ $deg\ (v) = l$ or $deg\ (v) = k$ for fixed values $l$ and $k$, $l \neq k$. A *subgraph H* of a graph $G$ is a graph whose vertices and edges are subsets of those of $G$. A *loop* is an edge that connects a vertex to itself. *Multiple edges* are two or more edges that are incident to the same two vertices.

In this survey a graph is undirected, without loops or multiple edges. *K-geodetic graphs* are defined as graphs in which each pair of nonadjacent vertices has at most $K$ paths of minimum length between them. Thus, a $K$-geodetic graph is *geodetic* when $K = 1$, *bigeodetic* when $K = 2$, and *trigeodetic* when $K = 3$. The minimum number of vertices whose deletion (implies also the deletion of the edges incident to the deleted vertices) disconnects $G$ is called *vertex connectivity* of a graph $G$. A graph $G$ is called *p-connected* if its vertex connectivity is equal to $p$. A *block* is a graph whose vertex connectivity is $p > 1$.

A *cover* of a graph $G$ is a set $\{G_1, G_2, ..., G_m\}$ of complete subgraphs of $G$ such that $G_1 \cup G_2 \cup ... \cup G_m = G$. A cover of $G$ is called a $\Theta$-*cover* if any two elements of the cover are edge-disjoint.

Let $G$ be a graph having vertices $v_1, v_2, ..., v_n$. Let $A = \{G_1, G_2, ..., G_m\}$ be a cover of $G$, where $V(G_i) = \left\{ v_{i_1}, v_{i_2}, ..., v_{j_i} \right\}$, $1 \leq i \leq m$. For each $i$, $1 \leq i \leq m$, take new vertices $v_{i_1 i}, v_{i_2 i}, ..., v_{j_i i}$ and construct a complete graph $K(G_i)$ on these vertices. Take $n$ new vertices $v_{10}, v_{20}, ..., v_{n0}$ and connect $v_{i_l i}$ to $v_{i_l 0}$ for $1 \leq l \leq j_i$, $1 \leq i \leq m$. The resulting graph is denoted $G^*(A)$.

Consider a $(b, n, r, k, \lambda)$-design on a set $S = \{x_1, x_2, ..., x_n\}$. Let $K_n$ be a complete graph with vertex set $\{x_1, x_2, ..., x_n\}$ and $G_i$ be a complete graph on vertex set of $B_i$, $1 \leq i \leq b$. Clearly, $A = \{G_1, G_2, ..., G_b\}$ is a cover of $K_n$. Construct graph $K_n^*(A)$ and denote it $K_n^*(r, k, \lambda)$. This is a $k$-connected, biregular block with degree sequence $(r, k)$. It has $n(r + 1)$ vertices and $nr(k + 1)/2$ edges.

**Theorem 4.** ([3]) *Let* $\mu = \max[\max(|B_i \cap B_j| : i, j = 1,...,b, i \neq j), \lambda]$. *Any pair of nonadjacent vertices of $K_n^*(r, k, \lambda)$ has at most $\mu$ distinct paths of minimum length between them. The diameter of $K_n^*(r, k, \lambda)$ is either 4 if $B_i \cap B_j \neq \emptyset$ for every $i, j, i \neq j$ or 5 if $B_i \cap B_j = \emptyset$ for some pair of distinct values $i, j$.* ☐

**Corollary 1.** ([3]) *If $(b, n, r, k, \lambda)$ is a symmetric block design, then in $K_n^*(r, k, \lambda)$ there are at most $\lambda$ paths of minimum length between each pair of vertices.* ☐

**Corollary 2.** ([3]) *If $(b, n, r, k, 1)$ is a block design, then $K_n^*(r, k, 1)$ is geodetic of diameter 4 or 5.* ☐

## Construction of K-Geodetic Blocks using Triple Systems

**Theorem 5.** ([3]) *When $n \equiv 1$ or $3 \pmod 6$, $n \geq 7$, there exists a 3-connected biregular geodetic block on $\frac{n(n+1)}{2}$ vertices having diameter 4 or 5 and degree sequence $\left( \frac{n-1}{2}, 3 \right)$.*

*Proof.* According to Theorem 1, when $\lambda = 1$, triple system $\left(\frac{\lambda n(n-1)}{6}, n, \frac{\lambda(n-1)}{2}, 3, \lambda\right)$ takes the form $\left(\frac{n(n-1)}{6}, n, \frac{n-1}{2}, 3, 1\right)$ and it exists for $n \equiv 1$ or $3 \pmod 6$, $n \geq 7$, on a set $S$. For the complete graph $G_i$ on vertices of $B_i$, $1 \leq i \leq n(n-1)/6$, graphs $G_1, ..., G_{n(n-1)/6}$ form a $\Theta$-cover of the complete graph $K_n$ on vertex set $S$. Construct graph $K_n*\left(\frac{n-1}{2}, 3, 1\right)$. This graph has $\frac{n(n+1)}{2}$ vertices. According to Corollary 2, this is a geodetic graph of diameter either 4 if $V(G_i) \cap V(G_j) \neq \emptyset$ for every $i, j, i \neq j$ or 5 if $V(G_i) \cap V(G_j) = \emptyset$ for some pair of distinct values $i, j$. Note that $K_n*\left(\frac{n-1}{2}, 3, 1\right)$ is a 3-connected biregular graph with degree sequence $\left(\frac{n-1}{2}, 3\right)$ for $n \geq 7$. $\qquad\square$

**Theorem 6.** ([1]) *For every $n \equiv 0$ or 1 (mod 3), $n \geq 4$, there exists a 3-connected biregular bigeodetic block on $n^2$ vertices with diameter 4 or 5 and degree sequence ($n$-1, 3).*

*Proof.* According to Theorem 1, when $\lambda = 2$, triple system $\left(\frac{\lambda n(n-1)}{6}, n, \frac{\lambda(n-1)}{2}, 3, \lambda\right)$ takes the form $\left(\frac{n(n-1)}{3}, n, \ n-1, 3, 2\right)$ and it exists for $n \equiv 0$ or 1 (mod 3), $n \geq 4$. Thus, taking $G_i$ to be a complete graph on vertices of $B_i$, $1 \leq i \leq n(n-1)/3$, graphs $G_1, ..., G_{n(n-1)/3}$ form a cover of the complete graph $K_n$ on vertex set $S$. Construct graph $K_n*(n$-1, 3, 2). This graph has $n^2$ vertices. According to Theorem 4, this is a bigeodetic graph of diameter either 4 if $B_i \cap B_j \neq \emptyset$ for every $i, j, i \neq j$ or 5 if $B_i \cap B_j = \emptyset$ for some pair of distinct values $i, j$. It is easy to observe that $K_n*(n$-1, 3, 2) has degree sequence ($n$-1, 3) and is 3-connected for $n \geq 4$. $\qquad\square$

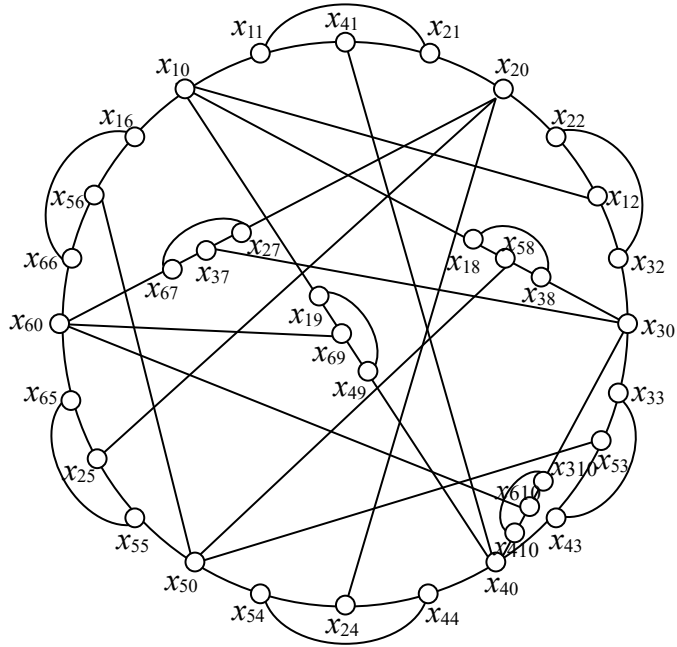

Fig. 1. A bigeodetic block generated by a (10, 6, 5, 3, 2)-design.

Next, we give the blocks of a (10, 6, 5, 3, 2)-design that are used to construct the bigeodetic block shown in Figure 1.

$$\{x_1, x_2, x_4\}, \ \{x_1, x_2, x_3\}, \ \{x_3, x_4, x_5\}, \ \{x_2, x_4, x_5\}, \ \{x_2, x_5, x_6\}, \ \{x_1, x_5, x_6\}, \ \{x_2, x_3, x_6\}, \ \{x_1, x_3, x_5\},$$
$$\{x_1, x_4, x_6\}, \ \{x_3, x_4, x_6\}.$$

**Theorem 7.** *For every $n \equiv 1 \pmod 2$, $n \geq 5$, there exists a 3-connected biregular trigeodetic block on $\frac{n(3n-1)}{2}$ vertices having diameter 4 or 5 and degree sequence $\left(\frac{3(n-1)}{2}, 3\right)$.*

*Proof.* According to Theorem 1, when $\lambda = 3$, triple system $\left(\frac{\lambda n(n-1)}{6}, n, \frac{\lambda(n-1)}{2}, 3, \lambda\right)$ takes the form $\left(\frac{n(n-1)}{2}, n, \frac{3(n-1)}{2}, 3, 3\right)$ and it exists for $n \equiv 1 \pmod 2$, $n \geq 5$. Thus, taking $G_i$ to be a complete graph on vertices of $B_i$, $1 \leq i \leq n(n-1)/2$, graphs $G_1, \ldots, G_{n(n-1)/2}$ form a cover of the complete graph $K_n$ on vertex set $S$. Construct graph $K_n^*\left(\frac{3(n-1)}{2}, 3, 3\right)$. This graph has $\frac{n(3n-1)}{2}$ vertices. According to Theorem 4, this is a trigeodetic graph of diameter either 4 if $B_i \cap B_j \neq \emptyset$ for every $i, j$, $i \neq j$ or 5 if $B_i \cap B_j = \emptyset$ for some pair of distinct values $i, j$. It is easy to observe that $K_n^*\left(\frac{3(n-1)}{2}, 3, 3\right)$ has degree sequence $\left(\frac{3(n-1)}{2}, 3\right)$ and is 3-connected for $n \geq 5$. □

### Construction of K-Geodetic Blocks using Symmetric Block Designs

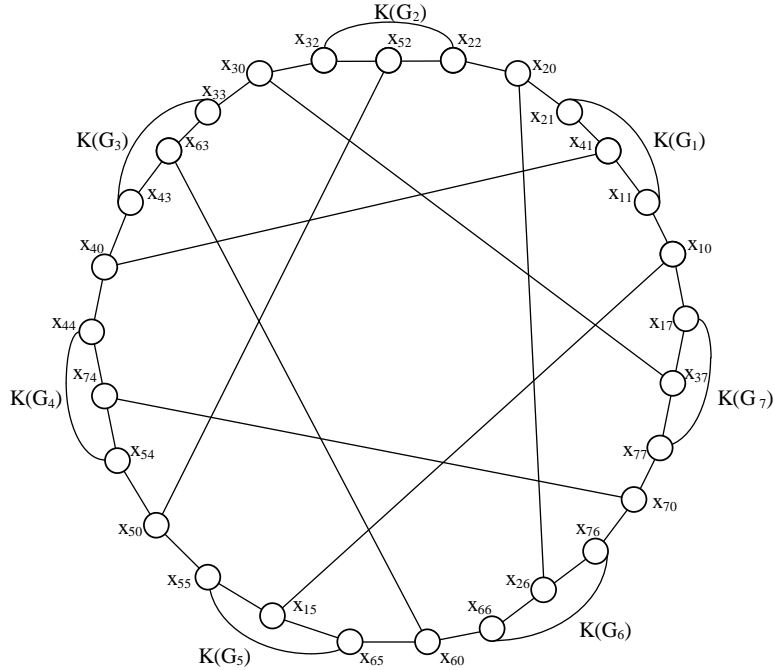

Fig. 2. A geodetic block generated by a (7, 7, 3, 3, 1)-design.

**Theorem 8.** ([3]) *When $n$ is a prime power, there exists an $(n+1)$-regular, $(n+1)$-connected geodetic block of diameter $4$.*

*Proof.* According to Theorem 3, when $n$ is a prime power, there exists a symmetric block design $(n^2+n+1,\ n+1, 1)$ on a set $S$ with blocks $B_i$, $1 \leq i \leq n^2+n+1$. Let $G$ be a complete graph on vertex set $S$, and $G_i$ be a complete graph on vertex set $B_i$, $1 \leq i \leq n^2+n+1$. $G_1$, ..., $G_{n^2+n+1}$ form a $\Theta$-cover of $G$. Construct graph $G^*_{n^2+n+1}(n+1,\ n+1,\ 1)$. This graph is an $(n+1)$-regular, $(n+1)$-connected one that has $(n^2+n+1)(n+2)$ vertices. According to Corollary 1, this is a geodetic block. Since any two blocks of a design $(n^2+n+1,\ n+1, 1)$ have one common element, the diameter of $G^*_{n^2+n+1}(n+1, n+1, 1)$ is 4. $\quad\square$

Next, we give the blocks of a $(7, 7, 3, 3, 1)$-design that are used to construct the geodetic block shown in Figure 2.

$$\{x_1, x_2, x_4\},\ \{x_2, x_3, x_5\},\ \{x_3, x_4, x_6\},\ \{x_4, x_5, x_7\},\ \{x_1, x_5, x_6\},\ \{x_2, x_6, x_7\},\ \{x_1, x_3, x_7\}.$$

**Theorem 9.** ([1]) *For every $n \equiv 1$ or $2 \pmod 4$, $2 \leq n \leq 10$, such that $(n\text{-}1)$ is a perfect square or $n \equiv 0$ or $3 \pmod 4$, $3 \leq n \leq 12$, such that $(n\text{-}1)$ is a prime power, there exists an $(n+1)$-regular, $(n+1)$-connected bigeodetic block of diameter $4$.*

*Proof.* According to Theorem 3, when $n \equiv 1$ or $2 \pmod 4$, $2 \leq n \leq 10$, such that $(n\text{-}1)$ is a perfect square or $n \equiv 0$ or $3 \pmod 4$, $3 \leq n \leq 12$, such that $(n\text{-}1)$ is a prime power, there exists a symmetric block design $((n^2+n+2)/2, n+1, 2)$ on a set $S$ with blocks $B_i$, $1 \leq i \leq (n^2+n+2)/2$. Let $G$ be a complete graph on vertex set $S$, and $G_i$ be a complete graph on vertex set $B_i$, $1 \leq i \leq (n^2+n+2)/2$. $G_1$, ..., $G_{(n^2+n+2)/2}$ form a cover of $G$. Construct graph $G^*_{(n^2+n+2)/2}(n+1, n+1, 2)$. This graph is an $(n+1)$-regular, $(n+1)$-connected one that has $(n^2+n+2)(n+2)/2$ vertices. According to Corollary 1, this is a bigeodetic block. Since any two blocks of a design $((n^2+n+2)/2, n+1, 2)$ have two common elements, the diameter of $G^*_{(n^2+n+2)/2}(n+1, n+1, 2)$ is 4. $\quad\square$

Next, we give the blocks of a $(7, 7, 4, 4, 2)$-design, which are used to construct the bigeodetic block shown in Figure 3.

$$\{x_1, x_2, x_3, x_4\},\ \{x_1, x_3, x_5, x_7\},\ \{x_1, x_4, x_5, x_6\},\ \{x_1, x_2, x_6, x_7\},\ \{x_2, x_3, x_5, x_6\},\ \{x_2, x_4, x_5, x_7\},$$
$$\{x_3, x_4, x_6, x_7\}.$$

**Theorem 10.** *For every $n \equiv 0$ or $2 \pmod 3$, $3 \leq n \leq 14$, $n \neq 12$, there exists an $(n+1)$-regular, $(n+1)$-connected trigeodetic block of diameter $4$.*

*Proof.* According to Theorem 3, when $n \equiv 0$ or $2 \pmod 3$, $3 \leq n \leq 14$, $n \neq 12$, there exists a symmetric block design $((n^2+n+3)/3, n+1, 3)$ on a set $S$ with blocks $B_i$, $1 \leq i \leq (n^2+n+3)/3$. Let $G$ be a complete graph on vertex set $S$, and $G_i$ be a complete graph on vertex set $B_i$, $1 \leq i \leq (n^2+n+3)/3$. $G_1$, ..., $G_{(n^2+n+3)/3}$ form a cover of $G$. Construct graph $G^*_{(n^2+n+3)/3}(n+1, n+1, 3)$. This graph is an $(n+1)$-regular, $(n+1)$-connected one that has $(n^2+n+3)(n+2)/3$ vertices. According to Corollary 1, this is a trigeodetic block. Since any two blocks of a design $((n^2+n+3)/3, n+1, 3)$ have three common elements, the diameter of $G^*_{(n^2+n+3)/3}(n+1, n+1, 3)$ is 4. $\quad\square$

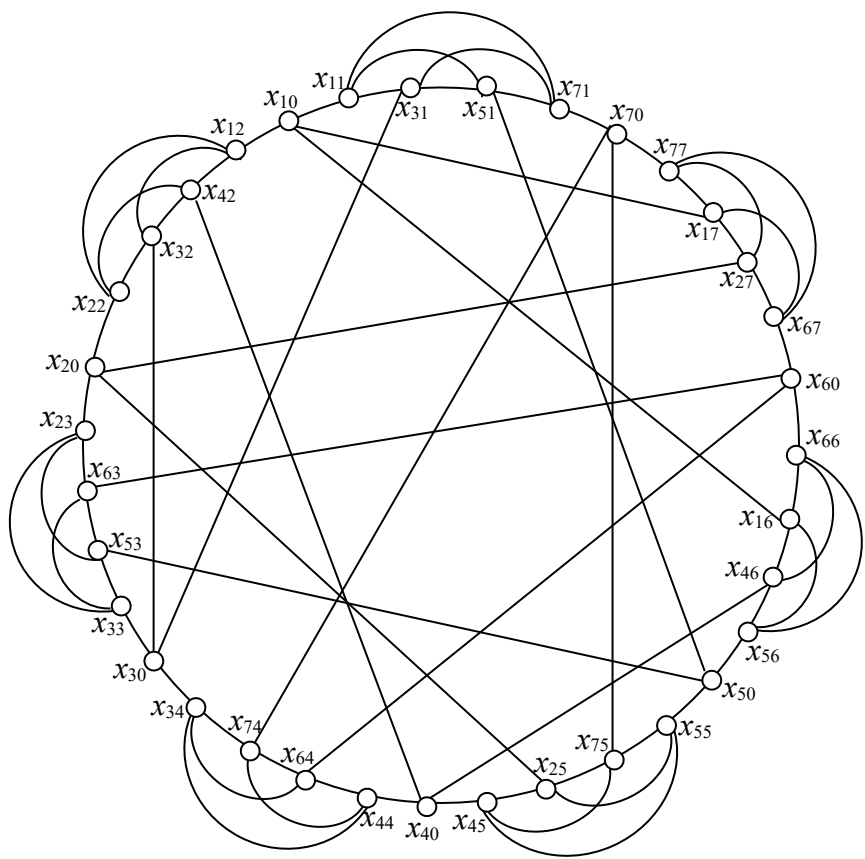

Fig. 3. A bigeodetic block generated by a (7, 7, 4, 4, 2)-design.